\documentclass[12pt]{article}
\usepackage{graphicx}
\usepackage{amsmath}
\usepackage{amssymb}
\usepackage{epsfig}
\usepackage{axodraw}
\usepackage{color}
\newcommand{\beeq}{\begin{equation}}
\newcommand{\eneq}{\end{equation}}
\newcommand{\beeqa}{\begin{eqnarray}}
\newcommand{\eneqa}{\end{eqnarray}}
\def\<{\langle}
\def\>{\rangle}

\begin{document}
\begin{titlepage}
\begin{flushright}
NSF-KITP-05-08, UPRF-2005-01 \\
\end{flushright}
\vspace{.4in}
\begin{center}
{\large{\bf  Exact lattice Ward-Takahashi identity for the $N=1$ Wess-Zumino model}  }
\bigskip \\ Marisa~Bonini$^{a}$ and Alessandra~Feo$^{a,b}$
\\
\vspace{\baselineskip}
{\small a. Dipartimento di Fisica, Universit\`a di Parma and INFN Gruppo Collegato di Parma, \\
Parco Area delle Scienze, 7/A,  43100 Parma, Italy \\
b. Kavli Institute for Theoretical Physics, UCSB, Santa Barbara, CA 93106, USA}
\mbox{} \\
\vspace{.5in}
{\bf Abstract} 
\bigskip 
\end{center} 
\setcounter{page}{0}
We consider a lattice formulation of the four dimensional $N=1$
Wess-Zumino model that uses the Ginsparg-Wilson relation. 
This formulation has an exact supersymmetry on the lattice. We show
that the corresponding Ward-Takahashi identity is satisfied, both at
fixed lattice spacing and in the continuum limit. The calculation is
performed in lattice perturbation theory up to order $g^2$ in the
coupling constant. We also show that this Ward-Takahashi identity
determines the finite part of the scalar and fermion renormalization
wave functions which automatically leads to restoration of
supersymmetry in the continuum limit. In particular, these wave
functions coincide in this limit.
\vspace{2.0cm}


\vspace{.5in}
\noindent
PACS numbers: 

\end{titlepage}
\section{Introduction}
Recently, there have been several attempts to study supersymmetric 
theories on the
lattice \cite{dondi}-\cite{curci} (for recent reviews and a
complete list of references, see \cite{review}).  The major obstacle
in formulating a supersymmetric theory on the lattice arises from the
fact that the supersymmetry algebra, which is actually an extension of
the Poincar\'e algebra, is explicitly broken by the space-time
discretization.  Without exact lattice supersymmetry one might hope to
construct non-supersymmetric lattice theories with a supersymmetric
continuum limit.  This is the case of the Wilson fermion approach for
the $N=1$ supersymmetric Yang-Mills theory \cite{curci} where the only operator
that violates the $N=1$ supersymmetry is a fermion mass term.  
By tuning the fermion mass to the supersymmetric limit one recovers
supersymmetry in the continuum limit (see Ref.~\cite{montvay} for numerical studies
along this approach).

The strategy of most recent studies is to realize part of the
supercharges as an exact symmetry on the lattice. This exact
supersymmetry is expected to play a key role to restore the continuum
supersymmetry without (or with less) fine-tuning of the action parameters. 
These ideas apply to theories with extended supersymmetry where the lattice theory
is realized by an orbifolding construction \cite{kaplan,rey}.
Another approach is based on writing the theory in terms of twisted fields 
\cite{catterall,sugino}. 
The connection between twisted fields and K\"ahler-Dirac fermions is emphasized in
\cite{dadda} and recently in \cite{catterall2}. 

In this paper we consider the $N=1$ four dimensional lattice
Wess-Zumino model introduced in Refs. \cite{fujikawa,fujikawa2} and
studied in \cite{bf} where it was shown that it is actually possible
to define a lattice supersymmetry transformation which leaves
invariant the full action at fixed lattice spacing.  This
transformation is non-linear in the scalar field.  The action and the
transformation are written in terms of the Ginsparg-Wilson operator
and reduce to their continuum expression in the naive continuum limit
$a \to 0$. In \cite{bf} the algebra of this lattice supersymmetry
transformation was studied and the closure of the algebra was
explicitly shown to $g^2$ order. This is a necessary ingredient
to guarantee the request of supersymmetry. It was also argued that
the existence of this exact symmetry is responsible for the
restoration of supersymmetry in the continuum limit.  In this paper,
we derive the Ward-Takahashi identity (WTi) associated with this
lattice supersymmetry transformation and show how in the continuum
limit one recovers the WTi associated with the continuum
supersymmetry transformation. This will be done in lattice
perturbation theory up to order $g^2$. An outcome of this approach is 
the calculation of the lattice renormalization wave function for the 
scalar and fermion fields.

The paper is organized as follows. In Sec.~\ref{wz} we briefly review
the $N=1$ four dimensional lattice Wess-Zumino model based on the
Ginsparg-Wilson fermion operator, and show how to build up a lattice
supersymmetry transformation which is an exact symmetry of the lattice
action. In Sec.~\ref{wtig} we derive the WTi and we explicitly check
the simplest one, the one-point WTi at one-loop. A second and
more interesting WTi, relating the boson and
fermion two-point function, is analyzed at $g^2$ order in Sec.~\ref{wtig2}. 
Here it is 
shown that this identity is exactly satisfied on the lattice.  
In Sec.~\ref{cont} we verify this WTi in the continuum limit and
determine the finite part of the lattice renormalization constants which allow
to identify the continuum invariant theory. 
Technical details of the fermion Dirac operator and the tadpole cancellations are presented in 
Appendix~A and B, respectively.

\section{The Wess-Zumino model}
\label{wz}
We formulate the lattice Wess-Zumino model by introducing a Dirac operator 
which satisfies the Ginsparg-Wilson relation \cite{ginsparg}
\beeq
\gamma_5 D + D \gamma_5 = a D \gamma_5 D\,.
\label{gw}
\eneq
This relation implies the existence of a continuum symmetry of the 
fermion action which may be regarded as a lattice form of the chiral 
symmetry \cite{luscher} and protects the fermion masses from additive 
renormalization.  As shown in Ref.~\cite{fujikawa2}, using this Dirac operator
it is possible to introduce a local action which is chiral invariant and where 
the fermions satisfy the Majorana condition. Moreover, in order to keep 
as much as symmetry as possible, the bosonic kinetic operator must be 
written in terms of $D$. 
The lattice action for the Wess-Zumino action reads
\beeq
S_{WZ} = S_0 + S_{int} \, , 
\label{wz2}
\eneq
with
\beeqa
&& S_{0} = \sum_x \bigg \{ \frac{1}{2} \bar \chi (1 - \frac{a}{2} D_1)^{-1} D_2 \chi - 
\frac{1}{a} ( A D_1 A + B D_1 B)  \nonumber \\ 
&& \phantom{S_{0} = \sum_x \bigg \{ }+ \frac{1}{2} F (1 - \frac{a}{2} D_1)^{-1} F + \frac{1}{2} G (1 - \frac{a}{2} D_1)^{-1} G 
\bigg\} \label{wz0} \, , \\[12 pt]
&& S_{int} = \sum_x \bigg\{ \frac{1}{2} m \bar \chi \chi + m (F A + G B) 
+ \frac{1}{\sqrt{2}} g \bar \chi (A + i \gamma_5 B) \chi \nonumber \\ 
&& \phantom{S_{int} = \sum_x \bigg \{ }+ \frac{1}{\sqrt{2}} g \big[ F (A^2 - B^2) + 2 G (A B) \big] \bigg\} \, ,
\label{wzint}
\eneqa
where $A$, $B$, $F$ and $G$ are real scalar fields and $\chi$ is a Majorana fermion
 which satisfies the Majorana condition 
\beeq
\bar \chi = \chi^T C
\eneq
\noindent
and $C$ is the charge conjugation matrix which satisfies 
\beeq
C^T = -C \, , \qquad \qquad C C^\dagger = 1 \, .
\eneq
Moreover, our conventions are 
\beeqa
&& C \gamma_\mu C^{-1} = - (\gamma_\mu)^T \, ,  \nonumber \\
&& C \gamma_5 C^{-1} = (\gamma_5)^T  \, .
\label{c}
\eneqa
The operators $D_1$ and $D_2$ which enter in $S_0$
are related to the operator $D$ in (\ref{gw}) by 
\beeq\label{d12}
D_1=\frac14\mbox{Tr}(D)\,, \;\;\;\;\;\;\;\;\;\;
D_2=\frac14\gamma_\mu\mbox{Tr}(\gamma_\mu D)\,.
\eneq
Our analysis is valid for all operators that satisfy Eq.~(\ref{gw}), however, in the following we will use the 
particularly simple solution given by Neuberger \cite{neuberger}
\beeq
D = \frac{1}{a} \bigg( 1 - X \frac{1}{\sqrt{X^\dagger X} } \bigg)\, , \qquad \qquad X = 1 - a D_w \, ,
\label{D}
\eneq
\noindent 
where
\beeq
D_w = \frac{1}{2} \gamma_\mu ( \nabla^\star_\mu + \nabla_\mu ) - \frac{a}{2} \nabla^\star_\mu \nabla_\mu 
\label{Dw}
\eneq
and 
\beeqa
\nabla_\mu \phi(x) &=& \frac{1}{a}(\phi(x + a \hat \mu) - \phi(x)) \, , \nonumber \\  
\nabla_\mu^\star \phi(x) &=& \frac{1}{a}(\phi(x) - \phi(x - a \hat \mu))   
\eneqa
are the forward and backward lattice derivatives, respectively.
\noindent
Substituting Eq.~(\ref{D}) in Eq.~(\ref{d12}) one finds
\beeq
D_1 = \frac{1}{a} \bigg[ 1 - (1 + \frac{a^2}{2} \nabla^\star_\mu \nabla_\mu) \frac{1}{\sqrt{X^\dagger X}} 
\bigg] \, , \qquad \qquad 
D_2 = \frac{1}{2} \gamma_\mu (\nabla^{\star}_\mu + \nabla_\mu) \frac{1}{\sqrt{X^\dagger X}} \equiv  
\gamma_\mu D_{2 \mu} \, .
\label{d1d2}
\eneq
The Ginsparg-Wilson relation (\ref{gw}) implies the following relations 
for $D_1$ and $D_2$ 
\beeq
D_1^2 - D_2^2 = \frac{2}{a} D_1 
\label{gw1}
\eneq
and 
\beeq
(1-\frac a2 D_1)^{-1}D_2^2=-\frac 2a D_1 \, .
\label{gw2}
\eneq

Before concluding this section we list the propagators of the 
lattice perturbation theory for the scalar and fermion fields: 
\beeqa\label{prop}
\< A A \> &=& \< B B\> = -{\cal M}^{-1} (1 - \frac{a}{2} D_1)^{-1}  \nonumber \\
\< F F\>&=& \< G G \> = \frac{2}{a}{\cal M}^{-1}  D_1  =-
{\cal M}^{-1} (1 - \frac{a}{2} D_1)^{-1} D_2^2  
\nonumber \\
\< A F \> &=& \< B G \> = m \,{\cal M}^{-1}  \nonumber \\
\< \chi \bar \chi \> & =& 
((1 - \frac{a}{2} D_1)^{-1} D_2  + m)^{-1}=
-{\cal M}^{-1} ((1 - \frac{a}{2} D_1)^{-1} D_2 - m) \, ,
\eneqa
where  
\beeq
{\cal M} = \Big[ \frac{2}{a} D_1 (1 - \frac{a}{2} D_1)^{-1} + m^2 \Big] 
\label{M}
\eneq
and the Ginsparg-Wilson relation (\ref{gw2})
has been used to rewrite the auxiliary fields propagators.
Despite the appearance of the operator $(1-\frac{a}{2}D_1)^{-1}$, there are no would be doublers and the propagators are regular (see appendix~A).

\subsection{The supersymmetric transformation}
As discussed in \cite{fujikawa}, $S_0$ is invariant under a lattice 
supersymmetry transformation which is obtained from the continuum one 
by replacing the continuum derivative with the lattice derivative $D_{2\mu}$. 
On the contrary the interaction term 
$S_{int}$ breaks this symmetry  because of the 
failure of the Leibniz rule at finite lattice spacing \cite{dondi}. 
In order to discuss the symmetry properties of the lattice Wess-Zumino model one possibility is to modify the 
action by adding irrelevant terms which make invariant the full action.
Alternatively, one can modify the supersymmetry transformation in such a way that the action 
(\ref{wz2}) has an exact symmetry for fixed $a$. 
In \cite{bf} it has been shown that the full action (\ref{wz2}) is invariant under 
the following supersymmetry transformation
\beeqa
&& \delta A = \bar \varepsilon \chi = \bar \chi \varepsilon  \nonumber \\
&& \delta B = -i \bar \varepsilon \gamma_5 \chi = -i \bar \chi \gamma_5 \varepsilon \nonumber \\
&& \delta \chi = - D_2 (A - i \gamma_5 B) \varepsilon - (F - i \gamma_5 G) \varepsilon + 
g R \varepsilon \nonumber \\ 
&& \delta F = \bar \varepsilon D_2 \chi \nonumber \\
&& \delta G = i \bar \varepsilon D_2 \gamma_5 \chi \, ,
\label{complete}
\eneqa
where $R$ is a function depending on the scalar fields
and their derivatives that can be determined in perturbation theory
imposing the invariance of the Wess-Zumino action under (\ref{complete}). 

By expanding $R$ in powers of $g$, 
\beeq
R = R^{(1)} + g R^{(2)} + \cdots
\label{expansion}
\eneq
and imposing the symmetry condition order by order in perturbation theory, one
finds 
\beeq
R^{(1)} = ((1 - \frac{a}{2} D_1)^{-1} D_2 + m )^{-1} \Delta L 
\label{r1}
\eneq
with 
\beeq
\Delta L \equiv  \frac{1}{\sqrt{2}} \Big\{2 (A D_2 A - B D_2 B) - D_2 (A^2 - B^2) 
+ 2 i \gamma_5 \Big[(A D_2 B + B D_2 A) - D_2 (A B)\Big] \Big\} 
\eneq
and
\beeq
R^{(n)} = - \sqrt{2} ((1 - \frac{a}{2} D_1)^{-1} D_2  + m)^{-1} (A + i \gamma_5 B)  R^{(n-1)}  \, ,
\label{rn}
\eneq
for $n \geq 2$.
Notice that the operator $((1 - \frac{a}{2} D_1)^{-1} D_2  + m)^{-1}$ is precisely the free fermion 
propagator and that the transformation (\ref{complete}), like the function $R$, 
is non-linear in the scalar fields. 
Indeed, using  (\ref{r1})  and (\ref{rn}) 
one sees that the expansion (\ref{expansion}) can be resummed and 
$R$ is the formal solution of the equation
\beeqa
&& \hspace{-0.8cm} \big[(1 - \frac{a}{2} D_1)^{-1} D_2  + m  + \sqrt{2} g (A + i \gamma_5 B) \big] R = 
\Delta L \, .  \label{Rnpt}
\eneqa
Notice that, in the limit $a \to 0$ the transformation (\ref{complete}) reduces
to the continuum supersymmetry transformation, since $\Delta L$
vanishes in this limit. Indeed, $\Delta L$ is different from zero
because of the breaking of the Leibniz rule at finite lattice
spacing.

In \cite{bf} it has been shown that the algebra associated with the lattice 
supersymmetry transformation (\ref{complete}) closes. The 
existence of this exact symmetry should be responsible for the restoration 
of supersymmetry in the continuum limit. 

In the following sections we will prove that the Ward-Takahashi identity (WTi)
derived from this lattice supersymmetry is exactly satisfied at finite lattice spacing. 
We will perform a one-loop analysis though the procedure can be generalized to higher loops.
We will also discuss the $a\to0$ limit.

\section{One-point Ward-Takahashi}
\label{wtig}
The WTi is derived from the generating functional
\beeq
Z[\Phi,J]=\int{\cal D}\Phi \exp{-(S_{WZ}+S_J)}
\eneq
where $S_J$ is the source term 
\beeq
S_J=\sum_x J_\Phi \cdot\Phi
\equiv \sum_x\Bigg\{J_A\,A+J_B\,B+J_F\,F+J_G\,G+\bar\eta\chi\Bigg\}\,.
\eneq
Using the invariance of both the Wess-Zumino action and the measure with respect
to the lattice supersymmetry transformation (\ref{complete}), the WTi reads
\beeq
\< J_\Phi \cdot \delta \Phi \>_J = 0\,,
\label{j}
\eneq
with $\delta\Phi$ given in (\ref{complete}).

We begin with the simplest WTi which is obtained by taking the derivative
with respect to $\bar\eta$ and setting to zero all the sources
\beeq\label{WT}
\<D_2(A - i \gamma_5 B)\> + \< F \> - i \gamma_5 \< G \> - g \< R \> = 0 \, .
\eneq
The order $O(g)$ of this Ward-Takahashi identity is
\beeq\label{wt1}
\<D_2(A - i \gamma_5 B)\>^{(1)} +\< F\>^{(1)} - i \gamma_5 \< G \>^{(1)} - g \< R^{(1)} \>^{(0)} = 0 \,,
\eneq
where the notation $\< {\cal O} \>^{(n)}$ indicates the $n-$order (in $g$) contribution to the expectation value 
of ${\cal O}$.
From Eq. (\ref{prop}) it is easy to see that all the terms of the WTi (\ref{wt1}) are zero. 
For instance
\beeq
\<D_2A\>^{(1)}\sim
D_{2xy} \Bigg[
\<A_yF_u\>\big[\<A_uA_u\>-\<B_uB_u\>\big]+
\<A_yA_u\>\big[2\<A_uF_u\>+2\<B_uG_u\>+\mbox{Tr}\<\bar\chi_u\chi_u\>\big]
\Bigg]
=0
\eneq
and similarly
\beeq
\<F_x\>^{(1)}\sim
\<F_xF_u\>\big[\<A_uA_u\>-\<B_uB_u\>\big]+
\<F_xA_u\>\big[2\<A_uF_u\>+2\<B_uG_u\>+\mbox{Tr}\<\bar\chi_u\chi_u\>\big]
=0 \, . \label{f1}
\eneq
The Feynman diagrams corresponding to the different contributions in (\ref{f1}) are depicted in
fig.~1. 

\begin{figure}[htbp]
\begin{center}
\includegraphics[height=2cm]{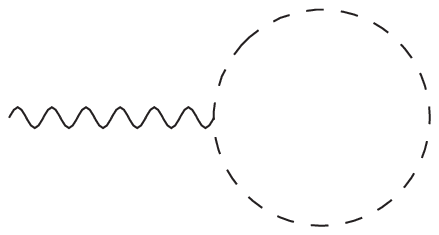}
\includegraphics[height=2cm]{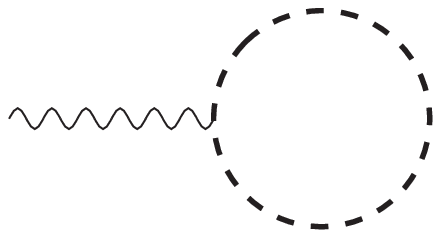}
\end{center}
\begin{center}
\includegraphics[height=2cm]{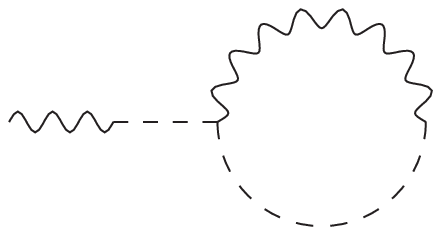}
\includegraphics[height=2cm]{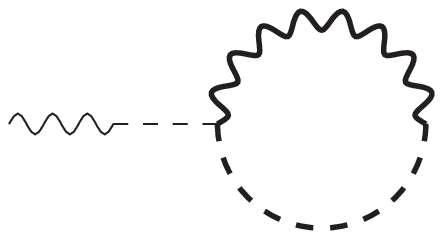}
\includegraphics[height=2cm]{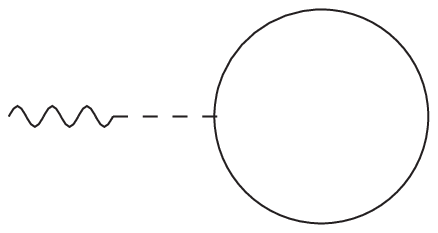}
\end{center}

%\begin{figure}[htbp]
%\begin{center}
%\epsfysize=2cm
%\epsfbox{tadaa.eps}
%\epsfysize=2cm
%\epsfbox{tadbb.eps}
%\end{center}
%\begin{center}
%\epsfysize=2cm
%\epsfbox{tadaf.eps}
%\epsfysize=2cm
%\epsfbox{tadbg.eps}
%\epsfysize=2cm
%\epsfbox{tadf.eps}
%\end{center}
\caption{\small{Tadpole cancellation.
The (bold) curly and 
(bold) dashed lines denote the auxiliary field ($G$) $F$  and the 
scalar field ($B$) $A$, respectively; the solid line denotes the 
fermion field.
}}
\end{figure}
The vanishing of the $A$ and $F$ one-point functions is due to the exact
cancellation of the tadpole diagrams on the lattice.  Similarly, the $G$ and $B$ 
one-point functions are zero at this order due to the
presence of a matrix $\gamma_5$ inserted in the fermion loop.
In order to prove the WTi (\ref{wt1}) one has to show that also the contribution depending on $R$ vanishes.
Indeed one finds
\beeqa
&&\< R^{(1)} \>^{(0)} =((1 - \frac{a}{2} D_1)^{-1} D_2  + m)^{-1}\<\Delta L_y\>^{(0)}
\nonumber\\
&&=
\<\chi_x\bar\chi_y\>\bigg[2\<A_yD_{2yz}A_z\>-2\<B_yD_{2yz}B_z\>-D_{2yz}\<A_zA_z\>+D_{2yz}
\<B_zB_z\>\bigg]
=0\,,
\eneqa
where (\ref{prop}) has been used.

\section{Two-point Ward-Takahashi identity}
\label{wtig2}
In this section we discuss a more interesting WTi that relates the 
fermion and scalar two-point functions.  Taking the
derivative of (\ref{j}) with respect to $\bar \eta$ and $J_A$ 
and setting to zero all the sources one obtains
\beeq
\<\chi_y\bar\chi_x\>-\<D_{2yz}(A_z - i \gamma_5 B_z)A_x\> - \<(F_y - i \gamma_5  G_y)A_x \> + g \< R_y A_x \> = 0 \, .
\label{WT22}
\eneq
Making use of the propagators given in (\ref{prop}), this identity is 
trivially satisfied at tree level. 

The next non-trivial order is $g^2$ which corresponds to the one-loop 
diagrams 
and can be written as
\beeq\label{WT2}
\<\chi_y\bar\chi_x\>^{(2)}
-\<D_{2yz}(A_z - i \gamma_5 B_z)A_x\>^{(2)} - \< (F_y - i \gamma_5  G_y)A_x \>^{(2)} + g \bigg(\< R^{(1)}_y A_x \>^{(1)} 
+g\< R^{(2)}_y A_x \>^{(0)} \bigg) = 0 \,,
\eneq
where we used the expansion (\ref{expansion}) for the function $R$.

Applying the Wick expansion, the first term of this WTi is 
\beeqa
&&\<\chi_y\bar\chi_x\>^{(2)}
=\frac{g^2}{4}
\<\chi_y\bar\chi_x\sum_{zu}
\Big[\bar \chi (A + i \gamma_5 B) \chi+ F (A^2 - B^2) + 2 G A B \Big]_z\nonumber\\
&& \phantom{<\chi_y\bar\chi_x>^{(2)}=\frac{g^2}{4}<\chi_y\bar\chi_x}
\times
\Big[\bar \chi (A + i \gamma_5 B) \chi+ F (A^2 - B^2) + 2 G A B \Big]_u
\>^{(0)}\,.
\eneqa
We first isolate among the various contributions the tadpole ones
\beeqa
&&\<\chi_y\bar\chi_x\>^{(2)}_T
=g^2\sum_{zu}\Big\{
\<\chi_y\bar\chi_z\>\<\chi_z\bar\chi_x\>\Big[
\<A_zF_u\>\Big(\<A_uA_u\>-\<B_uB_u\>\Big)
\nonumber\\
&&
\phantom{\<\chi_y\bar\chi_x\>^{(2)}_T=g^2\Big\{}
+ 2 \<A_zA_u\>\Big(\<A_uF_u\>+ \<B_uG_u\>\Big) 
-\<A_zA_u\>\mbox{Tr}\<\chi_u\bar\chi_u\>
\Big]
\nonumber\\
&&\phantom{\<\chi_y\bar\chi_x\>^{(2)}_T=g^2\Big\{}
+
\<\chi_y\bar\chi_z\>\gamma_5\<\chi_z\bar\chi_x\>\<B_zB_u\>\mbox{Tr}
\<\chi_u\gamma_5\bar\chi_u\>
\Big\}\,.
\eneqa
Using the propagators (\ref{prop}) and the relations
$\mbox{Tr}\<\chi\gamma_5\bar\chi\>=0$ and 
$\mbox{Tr}\<\chi\bar\chi\>=4\<AF\>=4\<GB\>$,
it is easy to demonstrate that the tadpole contributions cancel out (see Appendix~B).
This property is general and also holds for the other terms of the WTi
(\ref{WT2}). 
Therefore, one is left with  the connected non tadpoles diagrams
\beeq\label{wt21}
\<\chi_y\bar\chi_x\>^{(2)}_{NT}
=2g^2\sum_{uz}\Big\{
\<\chi_y\bar\chi_z\>\<\chi_z\bar\chi_u\>\<\chi_u\bar\chi_x\>\<A_zA_u\>-
\<\chi_y\bar\chi_z\>\gamma_5\<\chi_z\bar\chi_u\>\gamma_5\<\chi_u\bar\chi_x\>\<B_zB_u\>
\Big\} \, .
\eneq
The corresponding Feynman diagrams are given in fig.~2.\\
\begin{figure}[htbp]
\begin{center}
\epsfysize=1.8cm
\epsfbox{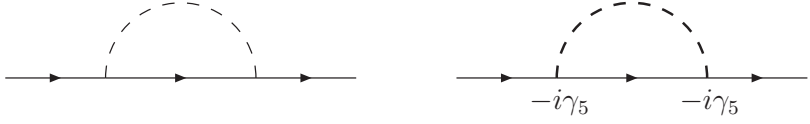}
\end{center}
\caption{\small{Non-tadpole contributions to $\<\chi\bar\chi\>^{(2)}$.
}}
\end{figure}

The non-tadpole contributions to the second term of (\ref{WT2}) are
(here and in the following the sum over repeated indices $z,u,w$ is understood)
\beeqa\label{wt22}
&&\<D_{2yz}(A_z - i \gamma_5 B_z)A_x\>^{(2)}_{NT}=g^2\Big\{
D_{2yz}\<A_zA_u\>\Big[\mbox{Tr}\Big(\<\chi_u\bar\chi_w\>\<\chi_w\bar\chi_u\> \Big)
+2\<A_uA_w\>\<F_uF_w\>
\nonumber\\&&
\phantom{\<D_{2yz}(A_z - i \gamma_5}
+2\<B_uB_w\>\<G_uG_w\>
+2\<F_uA_w\>\<A_uF_w\>+2 \<B_uG_w\>\<G_uB_w\>
\Big] \<A_wA_x\>
\nonumber\\&&
\phantom{\<D_{2yz}(A_z - i \gamma_5}
+D_{2yz}\<A_zF_u\>\Big[\<A_uA_w\>\<A_uA_w\>+\<B_uB_w\>\<B_uB_w\>\Big] \<F_wA_x\>
\nonumber\\&&
\phantom{\<D_{2yz}(A_z - i \gamma_5}
+2D_{2yz}\<A_zF_u\>\Big[\<A_uA_w\>\<A_uF_w\>-\<B_uB_w\>\<B_uG_w\>\Big] \<A_wA_x\>
\nonumber\\&&
\phantom{\<D_{2yz}(A_z - i \gamma_5}
+2D_{2yz}\<A_zA_u\>\Big[\<A_uA_w\>\<F_uA_w\>-\<B_uB_w\>\<G_uB_w\>\Big] \<F_wA_x\>\Big\} \, .
\eneqa
The corresponding Feynman diagrams are given in fig.~3.

\begin{figure}[htbp]
\begin{center}
\epsfysize=9.4cm
\epsfbox{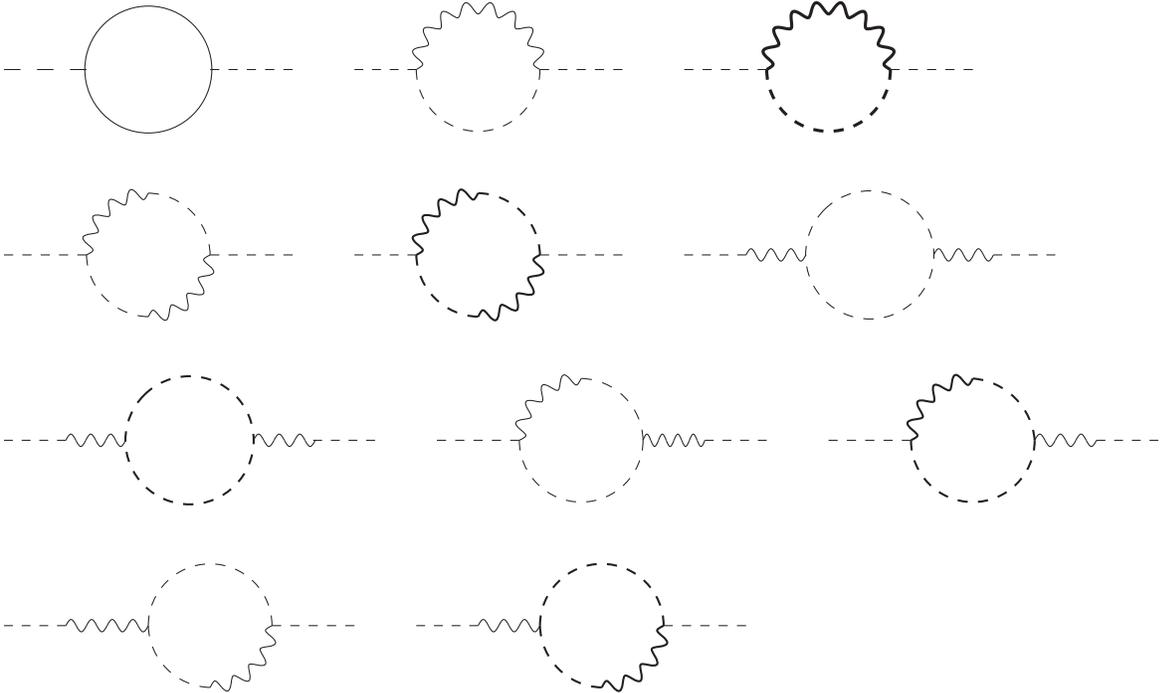}
\end{center}
\caption{\small{Non-tadpole contributions to $\<D_{2}(A-i\gamma_5 B)A\>^{(2)}$.
}}
\end{figure}
The non-tadpole contributions to the third term of (\ref{WT2}) are
\beeqa\label{wt23}
&& \< (F_y - i \gamma_5 G_y) A_x \>^{(2)}_{NT}=  g^2\Big\{2\< F_y A_u \> \Big[ 
\frac12 \mbox{Tr} \Big(\<\chi_u\bar\chi_w\> \<\chi_w\bar\chi_u\> \Big) + 
\< F_u F_w \> \< A_u A_w \> \nonumber \\
&& \phantom{ \< (F_y - i \gamma_5 G_y) A_x \>^{(2)}_{NT} }
+ \< F_u A_w \> \< A_u F_w \> + \< G_u G_w \> \< B_u B_w \> + 
\< B_u G_w \> \< G_u B_w \> \Big] \< A_w A_x\> \nonumber \\ 
&& \phantom{ \< (F_y - i \gamma_5 G_y) A_x \>^{(2)}_{NT} }
+ \< F_y F_u \> \Big[ \< A_u A_w \> \< A_u A_w \> 
+ \< B_u B_w \> \< B_u B_w \> \Big] \< F_w A_x \>  \nonumber \\
&& \phantom{ \< (F_y - i \gamma_5 G_y) A_x \>^{(2)}_{NT} }
+ 2 \< F_y A_u \> \Big[ \< F_u A_w \> \< A_u A_w \> 
- \< G_u B_w \> \< B_u B_w \> \Big] \< F_w A_x \>  
\nonumber \\
&& \phantom{ \< (F_y - i \gamma_5 G_y) A_x \>^{(2)}_{NT} }
+ 2 \< F_y F_u \> \Big[ \< A_u A_w \> \< A_u F_w \> 
- \< B_u B_w \> \< B_u G_w \> \Big] \< A_w A_x\> \nonumber \\
&& \phantom{ \< (F_y - i \gamma_5 G_y) A_x \>^{(2)}_{NT} }
- \gamma_5 \< G_y B_w \> \mbox{Tr} \Big( \gamma_5 \< \bar \chi_w \chi_u \> 
\< \bar \chi_u \chi_w \> \Big)  \<A_u A_x \>\Big\}  
\eneqa
and the corresponding Feynman diagrams are given in fig.~4.
\begin{figure}[htbp]
\begin{center}
\epsfysize=9.4cm
\epsfbox{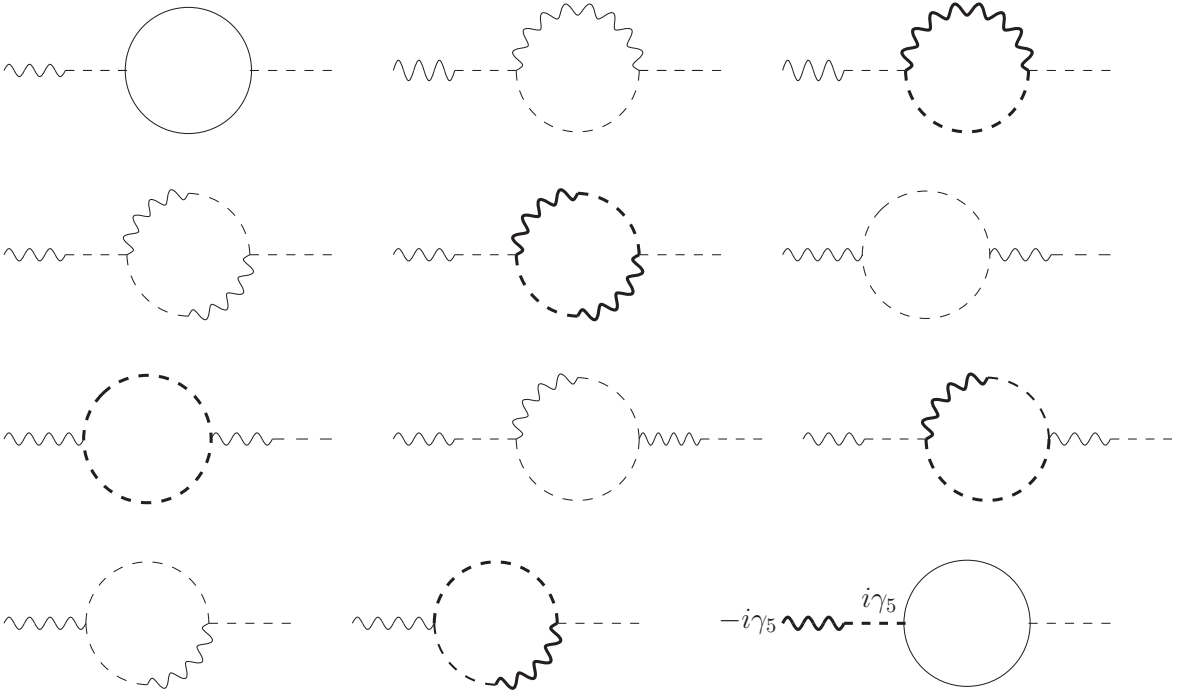}
\end{center}
\caption{\small{Non-tadpole contributions to $\< (F - i \gamma_5 G) A \>^{(2)}$.
}}
\end{figure}

Notice that the terms in last two rows of (\ref{wt22}) cancel out since $\<AA\>=\<B B\>$ and $\<AF\>=\<BG\>$. 
These terms, originating from the last four diagrams in fig.~3, are the one-loop contribution to the 1PI $AF$-vertex function, which therefore 
vanishes at this order. 
Similarly, the last three rows of (\ref{wt23}) do not contribute. In particular the last term,
i.e. the last diagram of fig.~4 vanishes and that gives $\< G_y A_x \>^{(2)}=0$. 

For the terms of the WTi (\ref{WT2}) involving the function $R$ one finds
\beeq
 \< R_y^{(1)} A_x \>^{(1)} = - \frac{g}{\sqrt{2}} \<\chi\bar\chi\>_{yz} 
\< \Delta L_z \big[
\bar \chi (A + i \gamma_5 B) \chi + F(A^2 - B^2) + 2 GAB \big]_u A_x \>^{(0)}
\,,
\eneq
where the fermion propagator follows from (\ref{r1}).

Also in this case the tadpole diagrams cancel out and one is left with 
\beeqa\label{r1a}
&& \hspace{-0.7cm}
\< R_y^{(1)} A_x \>^{(1)}_{NT} = -g\<\chi\bar\chi\>_{yz} \nonumber\\
&&\hspace{-0.7cm} \phantom{ \< R_y^{(1)} A_x\>^{(1)}_{NT} } 
\times\Big\{2 
 \Big[  \<A_z F_w\> D_{2zu} \< A_u A_w\> + \<A_z A_w\> D_{2zu} \< A_u F_w\> 
-D_{2zu}  \<A_u F_w\> \< A_u A_w\> 
\nonumber \\
&&\hspace{-0.7cm} \phantom{ \< R_y^{(1)} A_x\>^{(1)}_{NT} } 
- \<B_z G_w\> D_{2zu} \< B_u B_w\> -  \<B_z B_w\> D_{2zu} \< B_u G_w\> 
+D_{2zu}  \<B_u B_w\> \< B_u G_w\> 
\Big] \< A_w A_x \> 
\nonumber \\ 
&& \hspace{-0.7cm} \phantom{ \< R_y^{(1)} A_x\>^{(1)}_{NT} } 
+ \Big[ 2 \<A_z A_w\> D_{2zu} \< A_u A_w\> 
-D_{2zu}  \<A_u A_w\> \< A_u A_w\> 
\nonumber \\ && 
 \hspace{-0.7cm} \phantom{ \< R_y^{(1)} A_x\>^{(1)}_{NT}+\Big[ } 
+ 2\<B_z B_w\> D_{2zu} \< B_u B_w\> 
-D_{2zu} \<B_u B_w\> \< B_u B_w\> 
\Big] \< F_w A_x \> 
\Big\}\,.
\eneqa
The corresponding Feynman diagrams are given in fig.~5.

\begin{figure}[htbp]
\begin{center}
\epsfysize=4.5cm
\epsfbox{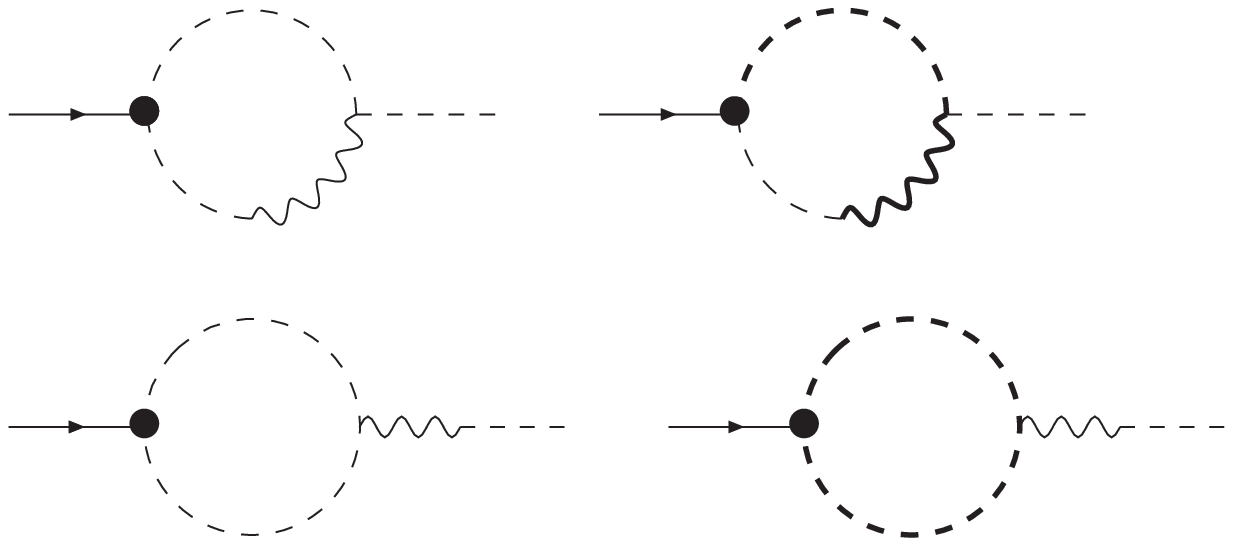}
\end{center}
\caption{\small{Non-tadpole contributions to $\< R^{(1)} A\>^{(1)}$. The blob 
denotes the insertion of the operator $D_2$ acting on the three legs outgoing from the vertex as in equation 
(\ref{r1a}).
}}
\end{figure}

Finally, for the last term of (\ref{WT2}) one gets
\beeqa
&&\< R_y^{(2)} A_x \>^{(0)} =
- \sqrt{2}\<\chi\bar\chi\>_{yz} 
\< (A_z+i\gamma_5B_z) \<\chi\bar\chi\>_{zw} \Delta L_w A_x\>^{(0)}
\nonumber\\ 
&&\phantom{\< R_y^{(2)} A_x \>^{(0)} }
= - 2\Big\{
\<\chi_y\bar\chi_z\> \<\chi_z\bar\chi_w\>\Big[\<A_zA_w\>D_{2wu}\<A_uA_x\>
+\<A_wA_x\>D_{2wu}\<A_zA_u\>
\nonumber\\ 
&&\phantom{\< R_y^{(2)} A_x \>^{(0)} =- 2\sqrt{2}\Big\{
\<\chi_y\bar\chi_z\> \<\chi_z\bar\chi_w\>}
-D_{2wu}\<A_zA_u\>\<A_uA_x\>\Big]
\nonumber\\ &&
\phantom{\< R_y^{(2)} A_x \>^{(0)}= }
-\<\chi_y\bar\chi_z\> \gamma_5\<\chi_z\bar\chi_w\> \gamma_5\Big[\<B_zB_w\>D_{2wu}\<A_uA_x\>
+\<A_wA_x\>D_{2wu}\<B_zB_u\>
\nonumber\\ 
&&\phantom{\< R_y^{(2)} A_x \>^{(0)} =- 2\sqrt{2}\Big\{
\<\chi_y\bar\chi_z\> \<\chi_z\bar\chi_w\>}
-D_{2wu}\<B_zB_u\>\<A_uA_x\>\Big] \Big\} \, ,
\label{r2}
\eneqa
and the corresponding Feynman diagrams are presented in fig.~6.

\begin{figure}[htbp]
\begin{center}
\epsfysize=2cm
\epsfbox{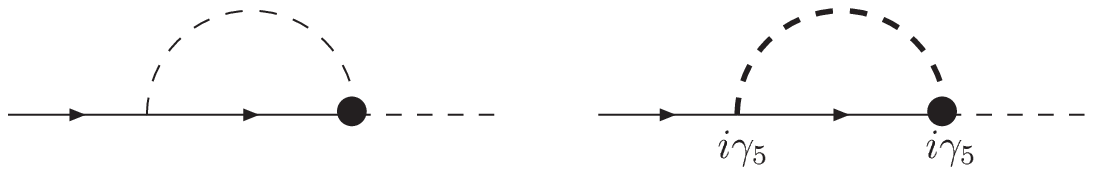}
\end{center}
\caption{\small{Non-tadpole contributions to $\< R^{(2)} A\>^{(0)}$. The blob 
denotes the insertion of the operator $D_2$ acting on the three legs outgoing from the vertex as in equation 
(\ref{r2}). 
}}
\end{figure}

\subsection{Calculation in momentum space}
In order to verify the WTi (\ref{WT2}) we find convenient to work in the momentum space representation.

For the fermion two-point function, 
the sum of the two diagrams in (\ref{wt21}) gives
\beeqa\label{fermion}
&&
\<\chi(p)\bar\chi(q)\>^{(2)}
=4g^2(2\pi)^4\delta^4(p+q)\bigg(D_2(p)-m (1 - \frac{a}{2} D_1(p))\bigg)
\int_k{\cal G}^{-1}(p,k) D_2(p+k) 
\nonumber\\
&&\phantom{\<\chi(p)\bar\chi(q)\>^{(2)}
=\frac{g^2}{4}(2\pi)^4\delta^4(p+q)}\times
\bigg(D_2(p)-m (1 - \frac{a}{2} D_1(p))\bigg)\,,
\eneqa
where
\beeq
{\cal G}(p,k)=\big[{\cal M}(p)((1 - \frac{a}{2} D_1(p))\big]^{2} 
\big[{\cal M}(k)((1 - \frac{a}{2} D_1(k))\big] 
\big[{\cal M}(k+p)((1 - \frac{a}{2} D_1(k+p))\big]
\label{gpk}
\eneq
and $D_1(p)$, $D_2(p)$ and ${\cal M}(p)$ are the Fourier transform of the operators given in (\ref{d1d2}) and (\ref{M}).

Similarly, the terms in (\ref{wt22}) and (\ref{wt23}) in momentum space write 
\beeqa\label{scalar}
&&\hspace{-1.4cm}
\<D_{2}(p) (A(p) - i \gamma_5 B(p))A(q)\>^{(2)}=
\< D_{2}(p) A(p)A(q)\>^{(2)}=g^2(2\pi)^4\delta^4(p+q)D_2(p)
\nonumber\\
&&
\phantom{D_{2}(p)}
\times\int_k {\cal G}^{-1}(p,k)
\Big[2m^2(1 - \frac{a}{2} D_1(p))^2
-\mbox{Tr}\big[D_2(k)D_2(p+k)\big]+4D_2^2(k)
\Big]
\eneqa
and
\beeqa\label{scalaux}
&&\hspace{-1cm}
\< (F(p) - i \gamma_5 G(p)) A(q)\>^{(2)}= 
\< F(p) A(q)\>^{(2)}= 
mg^2(2\pi)^4\delta^4(p+q)(1 - \frac{a}{2} D_1(p))
\nonumber\\
&&\hspace{-1cm}\phantom{ \< (F(p) - i \gamma_5 G(p))} 
\times\int_k {\cal G}^{-1}(p,k)
\Big[
\mbox{Tr}\bigg(D_2(k)D_2(p+k)\bigg)-4D_2^2(k)-2D_2^2(p)
\Big]\,,
\eneqa
respectively. 

Finally, the two terms in (\ref{r1a}) and (\ref{r2})
involving $R$ in momentum space become
\beeqa\label{erre1}
&& \< R^{(1)}(p) A(q)\>^{(1)} =  2 m g^2(2\pi)^4 \delta^4(p+q) 
(1 - \frac{a}{2} D_1(p)) 
\bigg(D_2(p) - m (1 - \frac{a}{2} D_1(p)) \bigg) 
\nonumber \\
&& \phantom{\< R^{(1)}(p) A(q)\>^{(1)} = } \times \int_k   
{\cal G}^{-1}(p,k) \bigg(  2 D_2(p+k) - D_2(p) \bigg)
\eneqa
and 
\beeqa \label{erre2}
&&  \< R^{(2)}(p) A(q)\>^{(0)} = -4g^2 (2\pi)^4 \delta^4(p+q)  \bigg(D_2(p) - m (1 - \frac{a}{2} D_1(p))\bigg) 
\nonumber \\ 
&& \phantom{ \< R^{(2)}(p) A(q)\>^{(0)} = } \times 
 \int_k  {\cal G}^{-1}(p,k)   D_2(p+k) 
\bigg(D_2(k) + D_2(p) - D_2(p+k)\bigg) \,.
\eneqa

In order to verify that the WTi (\ref{WT2}) is exactly satisfied, we find
convenient to arrange the various
terms according to the powers of $m$.

Inserting  (\ref{fermion}) and
(\ref{scalar})-(\ref{erre2}) into the WTi (\ref{WT2}) and setting $m=0$ one has
\beeqa
&& \hspace{-0.7cm} 4 g^2(2\pi)^4\delta^4(p+q)\int_k{\cal G}^{-1}(p,k) \Big[
D_2(p)D_2(p+k)D_2(p)+
 D_2(p) \Big(D_2(k) \cdot D_2(p+k) - D_2^2(k) \Big)
\nonumber \\ && \hspace{-0.7cm}  
\phantom{g^2(2\pi)^4\delta^4(p+q)\int_k{\cal G}^{-1}(p,k) \Big[  } 
-D_2(p)D_2(p+k) \Big(D_2(k) +D_2(p) - D_2(p+k)\Big)  \Big] \, ,
\eneqa
where $\mbox{Tr}(\gamma_\mu \gamma_\nu) = 4 \delta_{\mu \nu}$ has been used. 
Taking advantage of the invariance of ${\cal G}(p,k)$ under the change of variables $k \to -k -p$, 
one can replace $D_2(p+k) D_2(k)$  with $ \frac12 \left\{ D_2(p+k), D_2(k) \right\} = D_2(p+k) \cdot D_2(k) $
and therefore the integrand exactly vanishes.

The terms proportional to $m$ add up to 
\beeqa
&& \hspace{-0.6cm} g^2(2\pi)^4\delta^4(p+q) (1 - \frac{a}{2} D_1(p)) \int_k{\cal G}^{-1}(p,k) \Big[ 
-4 \Big(D_2(p) D_2(p+k) + D_2(p+k) D_2(p)\Big) \nonumber \\ 
&& \hspace{-2cm} \phantom{g^2(2\pi)^4\delta^4(p+q)} 
+ 4 D_2^2(k) + 2D_2^2(p) -4D_2(k) \cdot D_2(p+k)
+ D_2(p) \Big(4 D_2(p+k) - 2 D_2(p)\Big) \nonumber \\
&& \hspace{-2cm} \phantom{g^2(2\pi)^4\delta^4(p+q) }
+ 4 D_2(p+k) \Big(D_2(k) + D_2(p) - D_2(p+k)\Big) \Big] \, .
\label{daa}
\eneqa
Performing the substitution $D_2(p+k)D_2(k)\to D_2(p+k)\cdot D_2(k)$ 
as described above it is easy to check that (\ref{daa}) vanishes.

Finally, the contribution left is the one proportional to $m^2$, i.e.
\beeqa
&& \hspace{-0.6cm} g^2(2\pi)^4\delta^4(p+q) (1 - \frac{a}{2} D_1(p))^2 \int_k{\cal G}^{-1}(p,k) \Big[
4 D_2(p+k) - 2 D_2(p) \nonumber \\ 
&& \hspace{-0.6cm} \phantom{g^2(2\pi)^4\delta^4(p+q) (1 - \frac{a}{2} D_1(p))^2 \int_k{\cal G}^{-1}(p,k) } 
-2 (2 D_2(p+k) - D_2(p)) \Big] 
\eneqa
which is trivially zero. 

This end up our proof that the WTi (\ref{WT2}) is exactly satisfied at 
finite lattice spacing. 

\section{Continuum limit}\label{cont}
In this section we study the continuum limit of the WTi (\ref{WT2}) and 
discuss the restoration of the continuum supersymmetry in this limit.
This will clarify the mechanism of cancellation between the different terms in the WTi and the role of 
the operator $g \< R(p) A(q)\>^{(2)}$.

Following the notation of Ref.~\cite{KY}, the operator $D$ in (\ref{D}) can be written as
\beeq
D(p) =\bigg[ \frac{-i \sum_\mu \gamma_\mu \sin(p_\mu a)}{2 [\omega(p) + b(p)]} + \frac{a}{2}\bigg]^{-1}
\eneq
where
\beeq
\omega(p)=\frac{1}{a} \bigg[ \sum_\mu \sin^2(p_\mu a) + (a b(p))^2 \bigg]^{1/2}
\label{w}
\eneq
and 
\beeq
b(p) = \frac{1}{a} \bigg[ \sum_\mu 2 \sin^2(\frac{p_\mu a}{2}) - 1 \bigg]\,.
\label{b}
\eneq
With this notation, the operators $D_1$ and $D_2$ ($D = D_1 + D_2$) are
\beeq
D_1(p) = \frac{\omega(p) + b(p)}{a \omega(p)}
\eneq
and 
\beeq
D_2(p) = \frac{i}{a^2 \omega(p)} \sum_\mu \gamma_\mu \sin(p_\mu a)\,.
\label{d2}
\eneq
Similarly, 
\beeq
(1 - \frac{a}{2} D_1) = \frac{\omega - b}{2 \omega} 
\label{d1in}
\eneq
and 
\beeq
{\cal M}^{-1}  (1 - \frac{a}{2} D_1)^{-1} = 2 \omega a^2 \bigg[ 4 (\omega+b) + a^2 m^2 (\omega - b) \bigg]^{-1}\,.
\eneq

Each term in the WTi (\ref{WT2}) is a function of the external momenta $p$ and can be written as
\beeq
I(p) = \int \frac{d^4k}{(2 \pi)^4} F(k,p)
\label{int}
\eneq
where the integration momenta $k \in [-\frac{\pi}{a},\frac{\pi}{a}] $. 
If the integral (\ref{int}) is ultraviolet convergent, its continuum limit is obtained substituting  
the function $F(k,p)$ with its continuum
equivalent. Otherwise, if (\ref{int}) is divergent and contains only massive propagators so that $F(k,p)$ 
is finite for any set of exceptional momenta, one can use the lattice version of the BPHZ technique 
\cite{reisz} by writing 
\beeq
I(p) \equiv I^c(p) +\, I^l (p) \, ,
\eneq
where 
\beeq
I^c(p) = \int {d^4 k\over (2 \pi)^4}  
\bigg[ F(k,p) - \sum_{n=0}^{n_F} 
{1\over n!} \, p_{\mu_1} \ldots p_{\mu_n} 
\bigg( {\partial\over \partial p_{\mu_1} } \ldots
{\partial\over \partial p_{\mu_n}} F(k,p) \bigg)_{p=0} \bigg] \,,
\label{ic}
\eneq
\beeq
I^l(p) = \int {d^4 k\over (2 \pi)^4} \sum_{n=0}^{n_F} 
{1\over n!} \, p_{\mu_1} \ldots p_{\mu_n} 
\bigg( {\partial\over \partial p_{\mu_1} } \ldots
{\partial\over \partial p_{\mu_n}} F(k,p)\bigg)_{p=0}
\label{il}
\eneq
and $n_F$ is the degree of divergence of the diagram. 
$I^c(p)$ is ultraviolet finite and therefore its continuum limit can be taken.
All the effects of the lattice regularization
remain in $I^l(p)$, which is simply a polynomial in the 
external momenta with coefficients given by zero-momentum lattice integrals. 
In the following, we compute the lattice contributions of the Green 
functions entering in the WTi (\ref{WT2}).
Before doing this computation, we comment on their continuum part, 
such as $I^c(p)$, containing the subtracted integrand.
Since the subtraction makes the integrals UV finite, the order of 
the limit of zero lattice spacing and the momentum
integral can be interchanged. Applying this procedure to  
$\< R A \>$ one immediately recognizes that its continuum part vanishes, 
since the function $R$ vanishes 
for $a \to 0$. This is also clear due to the presence in (\ref{erre1}) and (\ref{erre2}) of the factor  
\footnote{ Actually, in (\ref{erre1}) one must first make the change of variables  $k \to -k -p$ to rewrite $2 D_2(p+k)$
as $D_2(p+k) - D_2(k)$. } $D_2(k) +D_2(p) - D_2(p+k)$ 
 that vanishes in this limit. 
For this reason one can restrict the analysis of the WTi to their lattice part.

For the fermion two point function (\ref{fermion}), one has to consider the following integral
\beeq
- \frac{i}{a} \int \frac{d^4k}{(2 \pi)^4} \frac{1}{[(\omega' + b')_k+
\frac{a^2 m^2}{4}(\omega' -b')_k] }
\frac{\omega'_{p+k}}{[(\omega' + b')_{p+k} + \frac{a^2 m^2}{4} (\omega' - b')_{p+k}]} 
\sum_\mu \gamma_\mu \sin(k_\mu) 
\label{chi}
\eneq
where $k$ has been rescaled to $k \to k/a$ and we have defined 
\beeq
\omega'_k\equiv a\omega(k/a) =  \bigg[  1 - 4 \sum_\mu \sin^4(\frac{k_\mu }{2}) + 
4 \bigg(\sum_\mu \sin^2(\frac{k_\mu }{2})\bigg)^2  \bigg]^{1/2}
\eneq
and 
\beeq
\omega'_{p+k}\equiv a\omega(p+k/a) =  \bigg[  1 - 4 \sum_\mu \sin^4(\frac{(k+ap)_\mu }{2}) + 
4 \bigg(\sum_\mu \sin^2(\frac{(k+ap)_\mu }{2})\bigg)^2  \bigg]^{1/2}\,.
\eneq
Similarly, $b'_k\equiv ab(k/a)$ and $b'_{p+k}\equiv ab(p+k/a)$ and their espressions can be easily read from (\ref{b}).

The factor $1/a$ in (\ref{chi}) 
implies a linear UV divergence of this integral which is cured by performing a Taylor expansion
in $pa$ up to the first derivative. The first term of the Taylor expansion of (\ref{chi}) is odd in $k$, thus is zero,
while the first derivative is
\beeq
\int \frac{d^4k}{(2 \pi)^4} \frac{-i\sum_\mu \gamma_\mu \sin(k_\mu) }{[(\omega' + b')_k + 
\frac{a^2 m^2}{4}(\omega' -b')_k] } 
\sum_\rho p_\rho \frac{\partial}{\partial p_\rho a} \bigg[ \frac{\omega'_{p+k}}{[(\omega' + b')_{p+k} + 
\frac{a^2 m^2}{4} (\omega' - b')_{p+k}]} \bigg]_{p=0}
\label{integral}
\eneq
with
\beeqa
&& \hspace{-1.4cm} \frac{\partial}{\partial p_\rho a} \bigg[ \frac{\omega'_{p+k}}{[(\omega' + b')_{p+k} + 
\frac{a^2 m^2}{4} (\omega' - b')_{p+k}]} \bigg]_{p=0} = 
\frac{1}{[(\omega' + b')_k + \frac{a^2 m^2}{4} (\omega' - b')_k]^2}
\nonumber \\
&& \times \bigg( \frac{2}{\omega'_k} 
\sum_{\mu\ne\rho} \sin^2(\frac{k_\mu}{2}) \sin(k_\rho) 
(2\sum_\nu \sin^2(\frac{k_\nu}{2}) - 1) -  \omega'_k \sin(k_\rho)
\bigg)
\label{derivative}
\eneqa
plus terms proportional to  $a^2m^2$ which do not contribute in the 
limit $a\to 0$. 
Notice that in the denominator the term proportional to $a^2m^2$ must 
be kept in order to ensure the IR finiteness of the integral, since 
$(\omega' + b')_k\sim k^2/2$ for $k\to 0$.
Indeed, by substituing this derivative in (\ref{integral}) one sees that
the contribution from the last term of (\ref{derivative}) produces
a $\log(a^2 m^2)$ divergence (for $a\to0$) originating 
from the $k \to 0$ integration region, while the remaining terms give rise to 
a finite integral.

Therefore, 
including the external leg factors, 
the fermion two point function can be written as~\footnote{From now on the factor $(2 \pi)^4 \delta^{(4)}(p+q)$ 
will be understood.}
\beeq
\<\chi \bar\chi \>^{(2)}(p)
= \frac{(i \not p - m)}{(p^2 + m^2)} C_2 i \not p \frac{(i \not p - m)}{(p^2 + m^2)}
\label{chicont}
\eneq
where 
\beeq
C_2 =g^2  \int \frac{d^4k}{(2 \pi)^4} \frac{\omega'_k \sin^2(k_\rho)}{[(\omega + b)'_k +\frac{ a^2 m^2}{4} 
(\omega - b)'_{k}]^3} + C_{2 f}  
\label{c2}
\eneq
and  $C_{2 f}$ is a finite constant that, for our purposes,  need not to be computed.

For the scalar two point function (\ref{scalar}) one has to calculate
the following integral
\beeqa
&&
\int \frac{d^4k}{(2 \pi)^4} \bigg\{ \frac{ m^2}{2} \frac{\omega'_k}{[(\omega' + b')_k + \frac{a^2 m^2}{4} (\omega' - b')_k]}
\frac{\omega'_{p+k}}{[(\omega' + b')_{p+k} + \frac{a^2 m^2}{4} (\omega' - b')_{p+k}]} \nonumber \\[5pt]
&&\phantom{\int \frac{d^4k}{(2 \pi)^4} \bigg\{}
-\frac{1}{a^2} \frac{1}{(\omega' + b')_k + \frac{a^2 m^2}{4} (\omega' - b')_k} 
\frac{\omega'_{p+k}}{(\omega' + b')_{p+k} + \frac{a^2 m^2}{4} (\omega' - b')_{p+k}} \nonumber\\[5pt]
&&\phantom{\int \frac{d^4k}{(2 \pi)^4} \bigg\{-\frac{1}{a^2}}
\times
\bigg(
\frac{\sum_{\mu}  \sin^2(k_\mu)}{\omega'_k}-
\frac{\sum_{\mu}  \sin(k_\mu)\sin(k_\mu+ap_\mu)}{\omega'_{p+k}}\bigg) \bigg\}\,.
\label{sviluppo}
\eneqa
The first term can be evaluated directly at $pa = 0$ while for the second
we need a Taylor expansion up to the second 
derivative in $pa$ due to the factor $1/a^2$.
We first concentrate on the latter term. It vanishes at $pa=0$ and  
moreover its first derivative is odd in $k$ and therefore also this term 
of the expansion vanishes. Thus the scalar two point 
function is given in terms of the following integral
\beeqa
&&\hspace{-1.2cm}
\int \frac{d^4k}{(2 \pi)^4} \bigg\{ \frac{ m^2}{2} 
\frac{(\omega'_k)^2}{[(\omega' + b')_k + \frac{a^2 m^2}{4} (\omega' - b')_k]^2}
\nonumber\\&&
-\frac{1}{(\omega' + b')_k + \frac{a^2 m^2}{4} (\omega' - b')_k}
\frac{1}{2} \sum_{\rho\sigma}p_\rho p_\sigma
\frac{\partial^2}{\partial p_\rho a\partial p_\sigma a}
\bigg[
\frac{\omega'_{p+k}}{(\omega' + b')_{p+k} + \frac{a^2 m^2}{4} (\omega' - b')_{p+k}} \nonumber\\[5pt]
&&\hspace{-1.2cm}
\phantom{\frac{1}{(\omega' + b')_k + \frac{a^2 m^2}{4} (\omega' - b')_k}}
\times
\bigg(
\frac{\sum_{\mu}  \sin^2(k_\mu)}{\omega'_k}-
\frac{\sum_{\mu}  \sin(k_\mu)\sin(k_\mu+ap_\mu)}{\omega'_{p+k}}\bigg) 
\bigg]_{p=0}\bigg\}\,.
\label{scalarcont}
\eneqa
There are two contributions coming from the second derivative. 
One is given by the product of 
(\ref{derivative}) with  
\beeqa
&&
\frac{\partial}{\partial p_\sigma a} 
\frac{\sum_{\mu}  \sin(k_\mu)\sin(k_\mu+ap_\mu)}{\omega'_{p+k}}
\bigg\vert_{p=0}
=\frac{\sin(k_\sigma)\cos(k_\sigma)}{\omega'_k}\nonumber\\
&&\phantom{\frac{\partial}{\partial p_\rho a} 
\frac{\sum_{\mu}  \sin(k_\mu)\sin(k_\mu+ap_\mu)}{\omega'_{p+k}}
\bigg\vert_{p=0}}
-\frac{2\sum_{\mu}\sin^2(k_\mu)\sum_{\nu\ne\sigma}\sin^2(\frac{k_\nu}{2})\sin(k_\sigma)}{(\omega'_{k})^3} \,,
\label{sviluppo2}
\eneqa
which produces a $\log(a^2 m^2)$ divergence to (\ref{scalarcont}),
originating from the product of the last term of (\ref{derivative})
with the first term of (\ref{sviluppo2}).  The second is given by the
second derivative of the third line of (\ref{scalarcont}) and its
explicit expression is not needed since its contribution to the
integral (\ref{scalarcont}) is finite for $a\to 0$.

Collecting all terms and including the external leg factors,  
the two point function (\ref{scalar}) becomes
\beeq
D_2 \< A  A \>^{(2)}(p) = 
i \not p \frac{1}{(p^2 + m^2)} \frac{1}{(p^2 + m^2)} (\frac12 C_3 m^2 - C_1 p^2)
\label{daacont}
\eneq
where 
\beeq
C_3 =g^2 \int \frac{d^4k}{(2 \pi)^4} \frac{(\omega'_k)^2}{[(\omega' + b')_k 
+ \frac{a^2 m^2}{4} (\omega' - b')_k]^2 } \,,
\label{c3}
\eneq
\beeq
C_1 =g^2\int \frac{d^4k}{(2 \pi)^4} \frac{\sin^2(k_\rho)\cos(k_\rho)}
{[(\omega' + b')_k + \frac{a^2 m^2}{4} (\omega' - b')_k]^3}+C_{1f} 
\label{c1}
\eneq
and $C_{1f}$ is a finite constant. 

A similar analysis applied to (\ref{scalaux}) and gives
\beeq
\< F A \>^{(2)}(p) = 
m \frac{1}{(p^2 + m^2)} \frac{1}{(p^2 + m^2)} (\frac12 C_3+C_1) p^2 \, .
\label{facont}
\eneq

The continuum limit of the two point function containing the operator $R$ can also be determined. For 
(\ref{erre1}) and (\ref{erre2}) one has
\beeq
\< R^{(1)} A \>^{(1)}(p) = m \frac{(i \not p - m)}{(p^2 + m^2)} \frac{1}{(p^2 + m^2)}
(C_2 -\frac12 C_3) i \not p
\label{erre1cont}
\eneq
and
\beeq
\< R^{(2)} A\>^{(0)}(p) = \frac{(i \not p - m)}{(p^2 + m^2)} \frac{1}{(p^2 + m^2)} (C_2 - C_1)p^2 \,,
\label{erre2cont}
\eneq
respectively.  Notice that the combinations $C_2-C_1$ and $C_2-\frac12
C_3$ are two (different) finite numbers.  Indeed, looking at the
$k\to0$ behavior of the integrand of (\ref{c2}), (\ref{c3}) and
(\ref{c1}) one sees that the $\log(a^2m^2)$ contributions cancels out
in these combinations. This is a consequence of the fact that the
one-loop correction to the two-point functions of $A$, $F$ and $\chi$
have the same logarithmic divergent parts \cite{fujikawa3}.

Substituting (\ref{chicont}), (\ref{daacont}) and
(\ref{facont})-(\ref{erre2cont}) in (\ref{WT2}) one verifies this WTi
in the continuum limit:
\beeqa
&& \frac{(i \not p - m)}{(p^2 + m^2)} ( i \not p C_2) \frac{(i \not p - m)}{(p^2 + m^2)}  \nonumber \\
&& - \frac{i \not p }{(p^2 + m^2)} ( \frac12 m^2 C_3 - p^2 C_1) \frac{1 }{(p^2 + m^2)} \nonumber \\
&& - \frac{m}{(p^2 + m^2)} (C_1 + \frac12 C_3) p^2  \frac{1 }{(p^2 + m^2)} \nonumber \\
&& + \frac{(i \not p - m)}{(p^2 + m^2)} (i \not p m) (C_2 - \frac12 C_3) \frac{1 }{(p^2 + m^2)} \nonumber \\
&& + \frac{(i \not p - m)}{(p^2 + m^2)} (C_2 - C_1) p^2 \frac{1 }{(p^2 + m^2)}
=0 \, .
\label{sum}
\eneqa
Actually, this is a check of the results we have obtained for the
continuum limit of the two point functions, since we have proved that
this WTi holds for any $a$ and therefore must be verified also in the
limit $a\to 0$.  Notice that the term $\< R A \>$ in (\ref{WT22}) is
essential to recover the WTi (\ref{WT2}) also for $a\to 0$.

Let us clarify the role of the operator $R$.
Thanks to the exactness of WTi (\ref{WT2}) 
it is always possible to write the two point function $\<R A \>^{(2)}$ 
as a suitable combination of the other three two point functions involved 
in this WTi.  
In particular, in the continuum limit one can write
\beeqa
&& \<R A\>=
\frac{i\not p-m}{p^2+m^2}i\not p\delta_1\frac{i\not p-m}{p^2+m^2}
+i\not p \frac{1}{p^2+m^2}(\delta_2 p^2+\delta_3 m^2)\frac{1}{p^2+m^2} \nonumber \\
&& \phantom{\<R A\>=} -\frac{m}{p^2+m^2}(\delta_2-\delta_3) p^2 \frac{1}{p^2+m^2}
\eneqa
where, from (\ref{erre1cont}) and (\ref{erre2cont}),
\beeq
\delta_1=\frac12C_3 -C_2-\delta_3\,,\;\;\;\;\;\;\;
\delta_2=\frac12C_3-C_1-\delta_3\,,\;\;\;\;\;\;\;
\eneq
and the constant $\delta_3$ is arbitrary.
Then in the continumm limit one 
can rewrite the WTi (\ref{WT2}) as the supersymmetric continuum WTi
\beeq
\< \chi \bar \chi \>_{R}^{(2)} - i\not p \< A A \>_{R}^{(2)} - \< F A \>_{R}^{(2)} = 0 
\eneq
with
\beeqa
&&\< \chi \bar \chi \>_{R}^{(2)} \equiv \< \chi \bar \chi \>^{(2)} +
\frac{i\not p-m}{p^2+m^2}i\not p\delta_1\frac{i\not p-m}{p^2+m^2}
\nonumber\\
&&
\< A A \>_{R}^{(2)} \equiv\< A A \>_{(2)}
-\frac{1}{p^2+m^2}(\delta_2 p^2+\delta_3 m^2)\frac{1}{p^2+m^2}
\nonumber\\
&&
\< F A \>_{R}^{(2)}\equiv \< F A \>^{(2)}+\frac{m}{p^2+m^2}
(\delta_2-\delta_3)p^2 \frac{1}{p^2+m^2}
\label{chinew}
\eneqa

It is convenient to express these two point functions in terms of
1PI vertex functions:
\beeqa
&&
\< \chi \bar \chi \>^{(2)}=
\frac{i\not p-m}{p^2+m^2}\,\Sigma^{(2)}_{\chi\bar\chi}\,\frac{i\not p-m}{p^2+m^2}\,,
\nonumber\\&&
\< A A\>^{(2)}=
-\frac{1}{p^2+m^2}(\Sigma^{(2)}_{AA}+m^2\Sigma^{(2)}_{FF})\frac{1}{p^2+m^2}\,,
\nonumber\\&&
\< F A\>^{(2)}=
\frac{1}{p^2+m^2}(\Sigma^{(2)}_{AA}-p^2\Sigma^{(2)}_{FF})\frac{m}{p^2+m^2}\,,
\eneqa
where the vanishing of the 1PI $AF$-vertex has been used. 
>From (\ref{chicont}), (\ref{daacont}) and (\ref{facont}), the lattice contribution to these 1PI vertices in 
the continuum limit reads
\beeq
\Sigma^{(2)}_{\chi\bar\chi}=i\not p  C_2\,,\;\;\;\;\;\;\;
\Sigma^{(2)}_{AA}=p^2C_1\,,\;\;\;\;\;\;\;
\Sigma^{(2)}_{FF}=-\frac12C_3\,.
\eneq
Moreover, from (\ref{chinew}), one has
\beeqa
&&
\Sigma^{(2)}_{\chi\bar\chi \,R}\equiv
\Sigma^{(2)}_{\chi\bar\chi}+i\not p\delta_1=i\not p(\frac{C_3}{2}-\delta_3)
\equiv- Z_{\chi}i\not p
\nonumber\\&&
\Sigma^{(2)}_{AA\,R}\equiv\Sigma^{(2)}_{AA}+p^2\delta_2=
p^2( \frac{C_3}{2}-\delta_3)\equiv- Z_{A}p^2
\nonumber\\&&
\Sigma^{(2)}_{FF\,R}\equiv\Sigma^{(2)}_{FF}+\delta_3=
-( \frac{C_3}{2}-\delta_3)\equiv Z_{F}\label{vertexnew}
\eneqa
with
\beeq
Z_{\chi}=Z_{A}=Z_{F}=-(\frac{C_3}{2}-\delta_3)\,.
\eneq
In Ref.~\cite{fujikawa3} it was shown that the one-loop corrections to the two-point function of $A$, $F$ and $\chi$ differ by finite quantities. Our construction shows that if one redefines the 1PI vertices as in (\ref{vertexnew})
the wave function renormalization factors become equal.
This is an important consequence of the exact lattice supersymmetry 
we have introduced and of the WTi derived from this symmetry.
This automatically leads to restoration of supersymmetry in the continuum 
limit with equal renormalization wave function for the scalar and fermion fields.

In a more standard approach \cite{fujikawa,kikukawaWZ}
the function $R$ is not included in the lattice supersymmetry transformation. 
Since the action is not invariant under this transformation, the WTi contains  
a breaking term. From the $a\to0$ limit of this WTi one determines the counterterms nedeed to restore supersymmetry in the continuum limit. 
The central issue of our approch is that this possibility is guaranteed 
by the existence of an exact supersymmetry of the lattice action.

\section{Conclusions}
\label{con}
In this paper, starting from the $N=1$ four dimensional lattice
Wess-Zumino model that uses the Ginsparg-Wilson relation and keeps an
exact supersymmetry on the lattice, it is showed that the corresponding
Ward-Takahashi identity is satisfied, both at fixed lattice spacing
and in the continuum limit. 
This result crucially depends on the Ginsparg-Wilson properties of the 
operators involved in the lattice action. The calculation is performed
in lattice perturbation theory up to order $g^2$ in the coupling constant.  

It is also showed that the study of the continuum limit of this Ward-Takahashi 
identity determines the finite part of the scalar and fermion renormalization wave functions which
automatically leads to restoration of supersymmetry in the continuum
limit. In particular, these wave functions coincide in this limit. 

Although we limit our computation up to the order $g^2$, this order is
not trivial and the discussion is general and applies to higher orders
by following the procedure described in this work.

There are several issues that remain to be investigated. First of all
it will be interesting to perform numerical simulations of this model
to check non-perturbatively the WTi (\ref{WT22}). Furthermore, one of the
most important question is whether these ideas may be extended to
theories with a gauge symmetry.

\vspace{1cm}\noindent{\large\bf Acknowledgements} \\
A.F. would like to thank Mike~Creutz, John Kogut, Herbert Neuberger and Tilo Wettig for organizing the 
program ``Modern Challenges in Lattice Field Theory'' at KITP where part of this work was completed. 
This research was supported in part by the National Science Foundation under
Grant No. PHY99-07949. 

\section{Appendix A}
\label{loc}
The fermionic kinetic term in the action (\ref{wz2}), 
\beeq
(1 - \frac{a}{2} D_1)^{-1} D_2
\label{zeromode}
\eneq
needs a careful study at the border of the Brillouin zone.
To this end we set
\beeq
p_\mu = (0,0,0,\frac{\pi - \epsilon}{a}) \label{000pi}
\eneq 
and  study the limit $\epsilon\to 0$.
Using  Eqs.~(\ref{w}) and (\ref{b}) 
we have that
\beeq
b = \frac{1}{a}\big[1 - \frac{\epsilon^2}{2}+{\cal O}(\epsilon^4)\big]
\eneq
and 
\beeq
\omega = \frac{1}{a} \, . 
\eneq
Inserting these values in (\ref{d2}) and (\ref{d1in}) we find
\beeq
(1 - \frac{a}{2} D_1)^{-1} =  \frac{4}{\epsilon^2}
\big[1+{\cal O}(\epsilon^2)\big]
\eneq
and 
\beeq
D_2 = i \gamma_4 \frac{\epsilon}{a} \big[1+{\cal O}(\epsilon^2)\big]
\, .
\eneq
Finally, the fermionic kinetic operator (\ref{zeromode}) behaves as
\beeq
(1 - \frac{a}{2} D_1)^{-1} D_2 = 4 i \gamma_4 \frac{1}{a \epsilon} 
\big[1+{\cal O}(\epsilon^2)\big]
\eneq
thus in the limit $\epsilon \to 0$ with fixed $a$ the would be doubler
(\ref{000pi}) becomes (infinitely) massive.  Similarly, one can 
check that this value of the momentum do not generate a pole in the
bosonic propagators.
This analysis can be generalized to the other edges of the Brillouin zone.
For instance, if
\beeq
p_\mu = (0,0,\frac{\pi - \epsilon}{a},\frac{\pi - \epsilon}{a})
\eneq
we have 
\beeq
(1 - \frac{a}{2} D_1)^{-1} D_2 = i (\gamma_3 + \gamma_4) \frac{3}{a \epsilon} 
\eneq
which again becomes a massive mode when $\epsilon \to 0$ while $a$ is kept fixed.
It is easy to demostrate that all the rest of the would be zero modes behave in the same way. 

Notice that the fermion propagator in (\ref{prop}) can be rewritten as 
\beeq
\< \chi \bar \chi \>  = 
-(D_2 - m(1 - \frac{a}{2} D_1)) \,\Big[ \frac{2}{a} D_1  + m^2 (1 - \frac{a}{2} D_1)\Big]^{-1}
\eneq
which is clearly finite for $\epsilon\to 0$.

\section*{Appendix B}
\label{appendixa}
In this appendix we explicitly show that the tadpole contributions to 
the two point WTi (\ref{WT2}) cancel separately. For the $\<\bar\chi_y \chi_x\>^{(2)}$ 
two point function this has been already shown in Section 5.
\\
The tadpole contribution to $\<D_{2yz}(A_z - i \gamma_5 B_z)A_x\>$
is
\beeqa
&&\hspace{-0.4cm}
\<D_{2yz}(A_z - i \gamma_5 B_z)A_x\>^{(2)}_{T}=g^2\Big\{
D_{2yz}\<A_zA_u\>
\Big[\<F_uA_w\>\Big(2\<F_wA_w\>+2\<G_wB_w\>-\mbox{Tr}\<\chi_w\bar\chi_w\>
\Big)\nonumber\\&&\hspace{-0.4cm}
\phantom{
\<D_{2yz}(A_z - i \gamma_5 B_z)A_x\>^{(2)}_{T}=g^2
D_{2yz}\<A_zA_u\>\Big]
}
+\<F_uF_w\>\Big(\<A_wA_w\>-\<B_wB_w\>\Big)\Big] \<A_uA_x\>
\nonumber\\&&\hspace{-0.4cm}
\phantom{\<D_{2yz}(A_z - i \gamma_5 B_z)A_x\>^{(2)}_{T}}
+
D_{2yz}\<A_zF_u\>
\Big[\<A_uA_w\>\Big(2\<F_wA_w\>+2\<G_wB_w\>-\mbox{Tr}\<\chi_w\bar\chi_w\>
\Big)\nonumber\\&&\hspace{-0.4cm}
\phantom{
\<D_{2yz}(A_z - i \gamma_5 B_z)A_x\>^{(2)}_{T}=
D_{2yz}\<A_zA_u\>}
+\<A_uF_w\>\Big(\<A_wA_w\>-\<B_wB_w\>\Big)\Big] \<A_uA_x\>
\nonumber\\&&\hspace{-0.4cm}
\phantom{\<D_{2yz}(A_z - i \gamma_5 B_z)A_x\>^{(2)}_{T}}
+
D_{2yz}\<A_zA_u\>
\Big[\<A_uA_w\>\Big(2\<F_wA_w\>+2\<G_wB_w\>-\mbox{Tr}\<\chi_w\bar\chi_w\>
\Big)\nonumber\\&&\hspace{-0.4cm}
\phantom{\<D_{2yz}(A_z - i \gamma_5 B_z)A_x\>^{(2)}_{T}=
D_{2yz}\<A_zA_u\>}
+\<A_uF_w\>\Big(\<A_wA_w\>-\<B_wB_w\>\Big)\Big] \<F_uA_x\>\Big\}
 \, .
\eneqa
Similarly, the tadpole contribution to $ \< (F_y - i \gamma_5 G_y) A_x \>$ is
\begin{multline}
\< (F_y - i \gamma_5 G_y) A_x \>^{(2)}_{T}=  
\< F_y A_u \> \Big[ 
\<F_uA_w\>\Big(2\<F_wA_w\>+2\<G_wB_w\>-\mbox{Tr}\<\chi_w\bar\chi_w\>
\Big)\\
+ \<A_uF_w\>\Big(\<A_wA_w\>-\<B_wB_w\>\Big)\Big] \<A_uA_x\>\\
+
\< F_y F_u \> \<A_uA_w\>
\Big[ 
2\<F_wA_w\>+2\<G_wB_w\>-\mbox{Tr}\<\chi_w\bar\chi_w\>
\Big]
\<A_uA_x\>\,.
\end{multline}
It is easy to see that both expressions are exactly zero.

\end{document}